\documentclass[aps,physrev,reprint,groupedaddress,longbibliography,letterpaper,floatfix]{revtex4-2}

\usepackage{graphicx}
\usepackage{dcolumn}
\usepackage{amsmath}
\usepackage{amsfonts}
\usepackage{bm}
\usepackage{physics}
\usepackage[dvipsnames]{xcolor}
\usepackage{hyperref}
\hypersetup{
colorlinks=true,
linkcolor=blue,
citecolor=blue,
urlcolor=blue,
}
\newcommand{\kB}{k_\mathrm{B}}
\newcommand{\dop}{\mathrm{d}}
\newcommand{\Ucontrol}{U_\mathrm{C}}
\newcommand{\Uland}{U_\mathrm{L}}
\newcommand{\Lfp}{\mathcal{L}^\mathrm{FP}}
\DeclareMathOperator*{\argmin}{arg\,min}
\newcommand{\evsm}[1]{\langle #1 \rangle}

\newcommand{\Ssys}{S_\mathrm{sys}}
\newcommand{\timeavg}[1]{\overline{#1}}
\newcommand{\numstates}{\mathcal{N}}
\newcommand{\Var}{\mathrm{Var}}

\begin{document}

\title{Optimal Control of Periodic Nonequilibrium Mechanochemical Systems via Automatic Differentiation}
\author{W.\ Callum Wareham}
\email{callum\_wareham@sfu.ca}
\author{David A.\ Sivak}
\email{dsivak@sfu.ca}
\affiliation{Department of Physics, Simon Fraser University, Burnaby, British Columbia V5A 1S6, Canada} 
\date{\today}

\begin{abstract}
Biological molecular machines are mesoscopic systems that act repeatedly and periodically to perform important cellular tasks while contending with strong fluctuations and operating in an overdamped regime.
Optimal control theory is a tool that can be used to understand the design principles behind efficient operation of these machines; however, most studies on optimal control of classical mesoscopic systems have focused on control problems that do not repeat periodically.
Here, we automatically differentiate Fokker-Planck simulations to design efficient nonequilibrium control strategies for simple models of periodic molecular machines with and without explicit changes in the machine's chemical state.
The designed protocols and theoretical analysis provide insight into the design principles governing efficient driving in these nonequilibrium systems.
Designed control protocols should seek to reduce mechanical heat by rotating the entire angular probability distribution at a constant speed without changing its shape, and should reduce chemical heat by reducing the proportion of chemical transitions with large heat.
\end{abstract}

\maketitle

\section{Introduction}

Biological molecular machines are a paradigmatic example of mesoscopic systems that are well-suited to the application of stochastic thermodynamics~\cite{brownTheoryNonequilibriumFree2020}.
Cells deploy a multitude of such machines to carry out the tasks critical to the organism's survival.
These machines transduce energies between different forms (e.g., chemical to mechanical), often with high efficiency~\cite{yasudaF1ATPaseHighlyEfficient1998,toyabeThermodynamicEfficiencyMechanochemical2011,rondelezHighlyCoupledATP2005,brownTheoryNonequilibriumFree2020}.
Compared with macroscopic machines, molecular machines must contend with different design challenges; in particular, they operate at low Reynolds number (with no momentum to carry them through their cycle) and using floppy components (biological polymers).
Additionally, they achieve high efficiency while repeating their task quickly and driven out of equilibrium by strong gradients.

For mesoscale systems with a goal that can be achieved by inputting energy via some control parameters (e.g., molecular machines, or physical bit erasure), optimal control theory is a framework for determining the most energy-efficient control-parameter schedule (a control \emph{protocol}) that reaches that goal~\cite{blaberOptimalControlStochastic2023}.
Studying optimal control protocols is a strategy for understanding efficient design principles of these systems.
Finding exact optimal protocols is typically intractable, except for simple systems~\cite{blaberOptimalControlStochastic2023}, so approximate methods for slow~\cite{sivakThermodynamicMetricsOptimal2012} and fast~\cite{blaberStepsMinimizeDissipation2021} driving are often applied.
Recently, more progress has been made on connecting these approximations in the limit of moderate driving speeds~\cite{zhongLinearResponseEquivalence2024}.

Most progress in optimal control of mesoscopic systems has been focused on one-pass problems, which seek the most efficient protocol for transferring the system between two set endpoints, often with the assumption that the system starts at equilibrium and, after the protocol concludes, is allowed to relax to equilibrium.
However, molecular machines are effective because they can repeat a task periodically and rapidly.
Some machines---for instance, adenosine triphosphate (ATP) synthase---even act as rotary motors, with a cycle that can be understood via a single physically intuitive mechanical coordinate~\cite{kawaguchiNonequilibriumDissipationfreeTransport2014, guptaOptimalControlF1ATPase2022, warehamMultiparameterOptimalControl2025}.
When operating quickly, there is no guarantee that a molecular machine equilibrates at any point in its cycle.

In this paper, we extend the literature on optimal control by numerically calculating periodic minimum-work control protocols away from equilibrium for a simplified model of a rotary molecular motor.
Inspired by Ref.~\cite{engelOptimalControlNonequilibrium2023}, we develop custom Fokker-Planck simulations and compute the gradient of a protocol's energy cost with respect to the protocol itself, which is iteratively used to update towards an optimal protocol.
Our code's use of automatically differentiated Fokker-Planck simulations of an ensemble distinguishes it from the JAX-MD~\cite{schoenholzJAXMDFramework2021} Langevin simulations in Ref.~\cite{engelOptimalControlNonequilibrium2023}.
Due to their cost, automatically differentiated computational methods of this scale have been enabled by the same new, powerful hardware accelerators that have been used in increasing numbers to meet the growing demand of popular artificial-intelligence techniques.
From our empirical results we distill intuitive design principles that generalize and complement principles from linear-response theory~\cite{sivakEquilibriumMeasurementsNonequilibrium2012,luceroOptimalControlRotary2019} and other studies on one-pass control~\cite{blaberEfficientTwodimensionalControl2022,zhongLimitedcontrolOptimalProtocols2022,engelOptimalControlNonequilibrium2023,zhongLinearResponseEquivalence2024,sawchukGlobalThermodynamicManifold2026}.

We consider models of molecular machines that incorporate the chemical drive from, e.g., ATP hydrolysis either implicitly or via explicit switching between chemical states of the machine.
These chemical transitions yield an additional avenue for producing waste heat, beyond the mechanical heat due to angular relaxation.
With insight from optimal-transport theory, we reason that optimal control protocols should attempt to reduce the mechanical waste heat by rotating the angular probability distribution at a constant speed and without changing its shape.
This produces a mechanical heat rate that is uniform across the entire protocol, and predicts a lower bound on the mechanical heat production at a given driving speed.
Our designed protocols qualitatively pursue this design principle, with varying degrees of success.
We also study the chemical heat, which our designed protocols attempt to reduce by controlling the average cost of a chemical transition.

In Sec.~\ref{sec:theory}, we review the molecular-machine models that we consider and the stochastic thermodynamics of those models, define the optimal control problem that we seek to solve, and predict design principles that our numerical methods should follow.
In Sec.~\ref{sec:methods}, we describe our numerical method and define the approximate versions of the stochastic-thermodynamic quantities presented in Sec.~\ref{sec:theory}.
In Sec.~\ref{sec:results}, we present the designed protocols for two simple molecular-machine models and compare their performance to corresponding naive and linear-response-designed protocols.
Finally, in Sec.~\ref{sec:discussion}, we discuss our results in a broader context, relating the design principles our protocols exhibit to those predicted by linear-response theory. 

\begin{figure}[!ht]
    \centering
    \includegraphics[width=0.99\columnwidth]{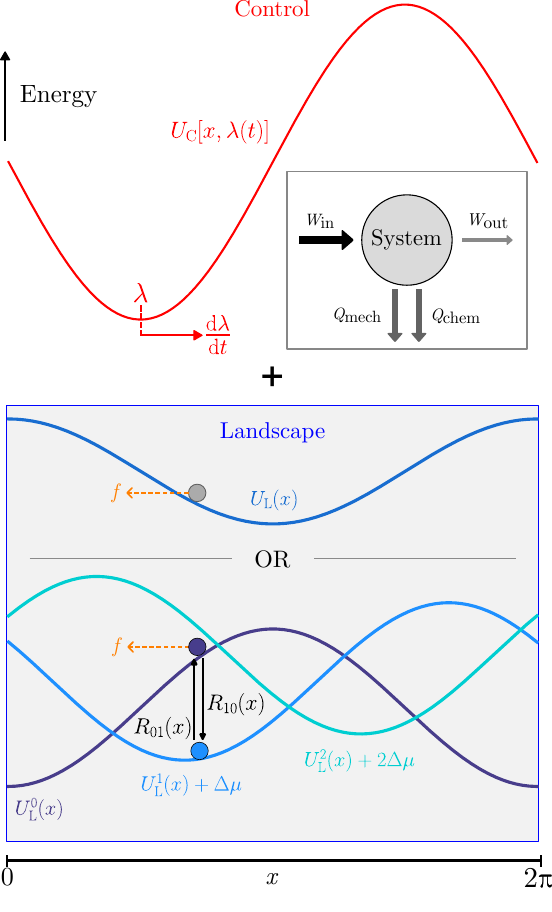}
    \caption{
        Schematic of the simple models for a molecular machine considered in this paper.
        We model the mechanical state of the molecular machine as a Brownian particle with angle $x$, which experiences a landscape potential $\Uland(x)$ (e.g., the blue curves) and a control potential $\Ucontrol[x,\lambda(t)]$ (red curve), the latter of which depends on time through the control parameter $\lambda(t)$ that evolves according to some control protocol $\Lambda$.
        The bead may experience a single constant landscape potential, or the potential may switch between a few landscapes according to chemical dynamics whose rates depend on the bead angle.
        Irrespective of the number of chemical states, we also allow for a constant nonconservative force $f$.
        Inset: Schematic of the energy flows to and from the system. The optimization procedure seeks to design a protocol that minimizes $\evsm{W_\mathrm{in}}$ for a given amount of output work $\ev{W_\mathrm{out}}$ (assumed fixed), which is equivalent to minimizing the total heat $\evsm{Q} = \evsm{Q_\mathrm{mech}} + \evsm{Q_\mathrm{chem}}$ produced by the driving protocol.
    }
    \label{fig:schematic}
\end{figure}

\section{Theory} \label{sec:theory}

\subsection{Model Systems} \label{sec:theory-models}

We model a rotary molecular motor with a simple but broadly applicable model.
We assume that the mechanical state of the molecular motor can be completely described by a single periodic coordinate, $x \in [0,2\pi)$, corresponding to (for example) the angular orientation of a rotating molecular axle.
We model the time evolution of $x$ as a one-dimensional Brownian particle diffusing subject to a time-dependent potential $U[x, \lambda(t)]$ and a constant nonequilibrium driving force $f$, with corresponding Langevin equation
\begin{equation}
    \dot{x} = \beta D\left(f-\frac{\partial U[x, \lambda(t)]}{\partial x}\right) + \sqrt{2 D} \, \eta(t) \ , \label{eq:langevin}
\end{equation}
where $\beta \equiv 1/\kB T$ is the inverse temperature, $D$ is the diffusion coefficient of the particle, and $\eta(t)$ is Gaussian white noise with zero mean and unit variance.
Since \eqref{eq:langevin} is often used to describe the probe bead in an experiment, throughout we refer to the object whose angle is $x$ as the \textit{bead}.  
We assume that the time-dependence of the internal energy enters only via the \textit{control parameter} $\lambda(t)$, which evolves according to a predetermined \textit{control protocol} $\Lambda$.
In this work, for simplicity, we restrict our attention to the case of a single control parameter, but the theory and results can be readily extended to account for multiple control parameters.

It will be useful to conceptualize the energy as having two components,
\begin{equation}
    U(x,t) = \Uland^{n(t)}(x) + \Ucontrol[x,\lambda(t)], \label{eq:gen_potential}
\end{equation}
where we identify the time-independent \textit{landscape potential} $\Uland$ that is not controlled, and the \textit{control potential} $\Ucontrol$ that is dynamically controlled.
It is useful to conceptualize $\Uland$ as the energy landscape resisting (or aiding) the `task' that the molecular machine seeks to accomplish, and $\Ucontrol$ as the machine's means of achieving that goal.
For some models, the landscape $\Uland$ switches between different potentials via the stochastic \textit{landscape state index} $n(t)$, which we constrain to positive integers up to the number $\numstates$ of landscapes.
The nonconservative driving force $f$ could stem directly from a constant physical torque and/or from coarse-graining over chemical states when chemical transitions are fast compared to dynamics~\cite{kawaguchiNonequilibriumDissipationfreeTransport2014,guptaOptimalControlF1ATPase2022,warehamMultiparameterOptimalControl2025}.

Figure~\ref{fig:schematic} is a schematic of the potential structures we consider.
For clarity, we delay discussion of the dynamics of $n$ to Sec.~\ref{sec:theory-switching}.
We will henceforth drop the explicit dependence of $\Ucontrol$ on the control parameter $\lambda(t)$ in favor of visually simpler $t$-dependence.

\subsubsection{Single Landscape}

We first consider the case of a single static landscape potential (i.e., $\numstates=1$). The bead's probability distribution $p_t(x)$ over the periodic angular coordinate (with $x + 2 \pi = x$) evolves according to the Fokker-Planck equation
\begin{subequations}
\begin{align}
    \frac{\partial p_t(x)}{\partial t} &= -\frac{\partial j_t(x)}{\partial x}, \label{eq:FPE_flux} \\
    &= \Lfp_t(x) \, p_t(x) \ , \label{eq:FPE_op} 
\end{align}
\end{subequations}
for \textit{probability flux}
\begin{equation}
    j_t(x) = D\left[\beta \left(  \frac{\partial U(x,t)}{\partial x} - f\right)p_t(x) -  \frac{\partial p_t(x)}{\partial x}\right] \ ,
\end{equation}
and the Fokker-Planck operator
\begin{equation}
    \Lfp_t(x) [\cdot] \equiv - D\frac{\partial}{\partial x} \left\{ \beta \left(\frac{\partial U(x,t)}{\partial x} -f \right)[\cdot] - \frac{\partial }{\partial x} [\cdot] \right\}. \label{eq:FP_op}
\end{equation}

We assume that the landscape potential has angular periodicity $\Uland(x + \xi) = \Uland(x)$, that the control protocol is time-periodic,
\begin{equation}
    \lambda(t+\tau)-\xi = \lambda(t) \implies \Ucontrol(x + \xi, t + \tau) = \Ucontrol(x,t) ,
\end{equation}
and the control potential has angular periodicity $\Ucontrol(x,t) = \Ucontrol(x + \chi, t)$.
Thus, the Fokker-Planck operator has periodicity
\begin{equation}
    \Lfp_{t+\tau}(x+ \xi) = \Lfp_{t}(x).
\end{equation}
Due to the $2 \pi$-periodicity of $x$ itself, we require that $2 \pi/\xi \in \mathbb{Z}^+$ and $2 \pi/\chi \in \mathbb{Z}^+$, i.e., the landscape and control periods are integer divisors of the angular-coordinate periodicity.

A useful coordinate transformation is to a frame of reference that rotates relative to the ``lab frame'' with velocity $\xi/\tau$ equal to the mean velocity of the control parameter,
\begin{equation}
    x' \equiv x - \xi t/\tau,
\end{equation} 
which we name the ``comoving frame''.
In these coordinates, $\Lfp$ has the form
\begin{multline}
    \Lfp_t(x')[\cdot] = \\- \frac{\partial}{\partial x'} \left\{ \left(-\frac{\xi}{\tau}+\beta D \left\{\frac{\partial U(x',t)}{\partial x'}-f\right\}\right) [\cdot] - D \frac{\partial }{\partial x'} [\cdot] \right\}, \label{eq:FP_comoving}
\end{multline}
with periodicity
\begin{equation}
    \Lfp_{t+\tau}(x') = \Lfp_{t}(x'). \label{eq:FP_space}
\end{equation}
The term $\xi/\tau$ in \eqref{eq:FP_comoving} arises from the force exerted on a bead at rest in the comoving frame by the apparently flowing fluid it is immersed in~\cite{speckRoleExternalFlow2008, seifertStochasticThermodynamicsFluctuation2012}.

In principle, Fokker-Planck equations obeying the structure of \eqref{eq:FP_space} can be solved via the \textit{Floquet ansatz}~\cite{jungPeriodicallyDrivenStochastic1993}, which asserts the solution structure
\begin{equation}
    p_t(x') = \sum_{\mu} c_\mu e^{-\mu t} p_t^\mu(x'),
\end{equation}
for constant weights $c_\mu$.
Here, the periodic Floquet eigenfunctions $p_t^\mu(x') = p_{t+\tau}^\mu(x')$ are solutions to the eigenproblem
\begin{equation}
    \left[\Lfp_t(x') - \frac{\partial}{\partial t}\right] \, p_t^\mu(x') = - \mu \, p_t^\mu(x') \label{eq:eigenproblem}
\end{equation}
corresponding to (complex) Floquet eigenvalues with $\Re(\mu) \geq 0$.
The full solution comprises a superposition of functions that repeat themselves each period $\tau$ (up to a constant factor $e^{-\mu \tau}$).  

After many consecutive repetitions of the protocol, such that $t \gg \tau$, only the $\mu=0$ term survives:
\begin{equation}
    p_t(x') \xrightarrow[t\gg \tau]{} c_0 \, p_t^0(x'). \label{eq:floquet_limit}
\end{equation}
Probability normalization requires that $1=\int \dop x \, c_0 \, p_t^0(x') = c_0$. The zeroth Floquet eigenfunction $p_0$ is therefore the initialization-independent \textit{periodic steady state} (PSS) for the periodic Fokker-Planck operator $\Lfp_t(x')$~\cite{jungPeriodicallyDrivenStochastic1993}, defined by $p_{t+\tau}(x')=p_t(x')$, or in the lab frame 
\begin{equation}
    p_{t+ \tau}(x+\xi) = p_t(x) \ .\label{eq:pss}
\end{equation}
Since effective molecular machines often repeat their cyclic tasks many times, it's likely that they spend most of their operational time in this periodic steady state---and the transient ``setup'' work by the control potential to approach that steady state is likely insignificant compared to the costs incurred from many repetitions in the steady state.
It is therefore natural to seek control protocols that are particularly efficient in the periodic steady state.

\subsubsection{Switching Landscapes} \label{sec:theory-switching}

Some rotary molecular motors (e.g., $\mathrm{F_1}$-ATPase) have a rotationally symmetric structure, and sequentially catalyze multiple chemical reactions through mechanical rotation~\cite{okunoRotationStructureFoF1ATP2011}.
Under the assumption that chemical reactions occur on timescales much faster than mechanical rotation, the machine's chemical state for each reaction contributes a (identically shaped, rotationally offset) mechanical landscape potential, with reactions (transitions between states) occurring stochastically at a rate dependent on the current rotational conformation.
For such cases, we model the system as having a joint probability distribution $p_t(x,n)$ over angle $x$ and landscape state $n$ at a particular time $t$.
We assume that each of the $\numstates$ landscape states are connected only with their direct neighbors (i.e., the only allowed transitions are $n \to n \pm 1$), and we identify the state indexed by $n=\numstates+1$ as state $n=1$, so that the state graph is an $\numstates$-cycle.

The probability distribution evolves via
\begin{multline}
    \frac{\partial}{\partial t} p_t(x,n) = \Lfp_t(x,n) \, p_t(x,n) \\
    + \sum_{m=n\pm 1} R_{nm}(x) \, p_t(x,m). \label{eq:FPE_switching}
\end{multline}
The state-dependent Fokker-Planck operator $\Lfp_t(x,n)$ is given by \eqref{eq:FP_op} with the potential structure of \eqref{eq:gen_potential}.
The angle-dependent transition rates $R_{nm}(x)$ for transitions $m \to n$ obey detailed balance,
\begin{equation}
    \frac{R_{n+1,n}(x)}{R_{n,n+1}(x)} = \exp \left\{\beta \, Q_\mathrm{chem}^{n+1, n}(x)\right\}, \label{eq:detbal}
\end{equation}
in terms of the chemical heat released during a single chemical transition $n\to n+1$,
\begin{equation}
    Q_\mathrm{chem}^{n+1, n}(x) \equiv  \Uland^{n}(x) - \Uland^{n+1}(x) - \Delta \mu_{n+1,n},
\end{equation}
and where $\Delta \mu_{n+1,n}$ is the chemical free-energy change associated with the transition $n\to n+1$ (e.g., the chemical-potential change associated with hydrolysis/synthesis of a fuel molecule).
The per-transition chemical heat is the difference in energy between the $(n+1)$th and $n$th landscape states, including the chemical free-energy change $\Delta \mu_{n+1,n}$.
Out of the many forms for the rates that are permitted by \eqref{eq:detbal}, we make a simple choice~\cite{kawaguchiNonequilibriumDissipationfreeTransport2014,guptaOptimalControlF1ATPase2022,warehamMultiparameterOptimalControl2025},
\begin{subequations} \label{eq:switching_rates}
\begin{align}
    R_{n+1,n}(x) &=\Gamma\, \exp \left\{ q \, \beta \, Q_\mathrm{chem}^{n+1, n}(x) \right\}, \\
    R_{n,n+1}(x) &= \Gamma  \, \exp \left\{ (q-1) \, \beta  \, Q_\mathrm{chem}^{n+1, n}(x) \right\}.
\end{align}
\end{subequations}
The \textit{switching-rate} parameter $\Gamma$ tunes the frequency at which chemical transitions occur; $\Gamma$ increases, for example, when the concentrations of reactant and product small molecules are raised at fixed chemical potential $\Delta \mu$~\cite{kawaguchiNonequilibriumDissipationfreeTransport2014,guptaOptimalControlF1ATPase2022,nakayamaAsymmetricEnzymeKinetics2025}.
The \textit{asymmetry parameter} $q \in [0,1]$ controls the degree to which each of the forward and backward rates depends on the per-transition chemical heat at each angle $x$.
The choice $q = 1/2$ corresponds to no asymmetry in the angle-dependence between the forward and backward rates, while the choice $q=1$ ($q=0$) yields a system where only the rate of transitions $n\to n+1$ ($n+1\to n$) depends on $x$, with the reverse $x$-independent.
$\mathrm{F}_1$-ATPase, for example, is thought to be fully asymmetric, with ATP hydrolysis rate independent of shaft angle~\cite{kawaguchiNonequilibriumDissipationfreeTransport2014,nakayamaAsymmetricEnzymeKinetics2025}.

Many rotary molecular motors are composed of a series of identical monomers~\cite{okunoRotationStructureFoF1ATP2011,guoBacterialFlagellarMotor2022,martinIondrivenRotaryMembrane2024}, so in this work we assume each landscape potential has the same shape and that they are equally spaced in $x$,
\begin{equation}
    \Uland^n(x \mp \xi) = \Uland^{n \pm 1}(x),~\mathrm{with}~\xi \equiv 2 \pi/\numstates,
\end{equation}
and that the free-energy difference between neighboring potentials is uniform, $\Delta \mu_{n+1,n} = - \Delta \mu_{n-1,n} = \Delta \mu$.
The right-hand side of \eqref{eq:FPE_switching} still satisfies the periodicity condition required for application of the Floquet ansatz, so steady states in the form of \eqref{eq:pss} will hold in the long-time limit for the joint distribution, $p_{t+\tau}(x+\xi,n) = p_t(x,n)$.

\subsection{Stochastic Thermodynamics}

Stochastic thermodynamics generalizes traditional thermodynamics to stochastic systems, identifying well-defined work, heat, and entropy flows~\cite{pelitiStochasticThermodynamics2021}. For probability distributions $p_t(x)$ over a continuous variable like the ones we study here, we identify a work rate $\evsm{\dot{W}}$ and a heat rate $\evsm{\dot{Q}}$ via the chain-rule expansion of the time derivative of the internal energy,
\begin{subequations}
\begin{equation}
    \evsm{\dot{E}} = \frac{\partial}{\partial t} \int \dop x \, p_t(x) \, U(x,t)
\end{equation}
\begin{multline}
    = \underbrace{\int \dop x \, p_t(x) \frac{\partial U(x,t)}{\partial t}}_{=\evsm{\dot{W}}} \\
    + \underbrace{\int \dop x \, \frac{\partial p_t(x)}{\partial t}\left[U(x,t) + fx\right]}_{=\evsm{\dot{Q}}}.\label{eq:heat_work_defn}
\end{multline}
\end{subequations}
The angle brackets $\evsm{\cdot}$ indicate an average over a (generally nonequilibrium) ensemble of stochastic bead trajectories, with each realization differing in initial condition and noise trajectory $\eta(t)$.
The standard decomposition~\eqref{eq:heat_work_defn} is useful in models with a single landscape, but in Sec.~\ref{sec:st_switching} we describe the modifications required for switching landscapes. 
In this section we define the energy flows [shown schematically in Fig.~\ref{fig:schematic} inset].

Work on the system is either dissipated as heat or retained to change the system energy,
\begin{equation}
    \evsm{W} = \evsm{Q} + \evsm{\Delta E} \ . \label{eq:nrg_bal}
\end{equation}
The PSS condition implies that over a complete cycle the system entropy is unchanged, thus any change in internal energy equals the change in (equilibrium) free energy, $\evsm{\Delta E} = \Delta F$.
We consider the free-energy change $\Delta F$ to be the ``useful'' work; the heat $\evsm{Q}$ is the energy lost in applying the control protocol.
Throughout this paper, we limit our analysis to potential structures and driving speeds that produce a vanishingly small proportion of trajectories that escape the local control well, so that each full trap rotation produces on average a full rotation of the bead, and the average output work $\evsm{W_{\rm out}}$ can be tracked by the control parameter $\lambda(t)$.
The optimization problem therefore amounts to minimizing $\evsm{Q}$.

The rate of entropy production in the medium is identified with the heat rate,
\begin{equation}
    \dot{S}_\mathrm{m} = \frac{\evsm{\dot{Q}}}{T}. \label{eq:ent_heat}
\end{equation}
The probability distribution itself supports an information-theoretic entropy
\begin{equation}
    \Ssys(t) = - \int \dop x \, p_t(x) \ln p_t(x).  
\end{equation}
The total entropy production rate of the universe is the sum of the medium and system entropy rates,
\begin{equation}
    \dot{S}_\mathrm{tot} = \dot{S}_\mathrm{m} + \dot{S}_\mathrm{sys},
\end{equation}
which satisfies the second law, $\dot{S}_\mathrm{tot} \geq 0$~\cite{seifertEntropyProductionStochastic2005,seifertStochasticThermodynamicsFluctuation2012}.

\subsubsection{Single Landscape} \label{sec:theory-stoch-single}

When $p_t(x)$ evolves on a single continuous landscape according to the Fokker-Planck equation~\eqref{eq:FPE_op}, the work done by the controller on the bead is
\begin{equation}
    \evsm{\dot{W}} = \int \dop x \, p_t(x) \frac{\partial \Ucontrol(x,t)}{\partial t}.
\end{equation}
The rate of free-energy change is~\footnote{Throughout, care has been taken to ensure that there are no artifacts in the calculation of expectation values of angle (e.g., in $\ev{x}$) that could arise from a poor choice of periodic dynamical reference frame.}
\begin{equation}
    \dot{F} =  f\int \dop x \, j_t(x).
\end{equation}
In the PSS, the free-energy change over a single period $\tau$ is
\begin{equation}
    \Delta F_\tau = \pm f \xi, \label{eq:free_nrg_single}
\end{equation}
where the sign depends on if the bead is moving against (+) or with (-) the nonequilibrium torque.

The total entropy production rate can be written as~\cite{seifertEntropyProductionStochastic2005,seifertStochasticThermodynamicsFluctuation2012,nakazatoGeometricalAspectsEntropy2021,chennakesavaluUnifiedGeometricFramework2023}
\begin{subequations}
\begin{align}
    \dot{S}_\mathrm{tot} &= \frac{\kB}{D}\int \dop x \, \frac{j_t(x)^2}{p_t(x)} \\
    &= \frac{\kB}{D} \evsm{\nu^2},
\end{align}
\end{subequations}
for the local mean velocity
\begin{subequations}
\begin{align}
    \nu_t(x) &\equiv \frac{j_t(x)}{p_t(x)} \label{eq:nu_def} \\
    &= \evsm{\dot{x}|x,t}. 
\end{align}
\end{subequations}
The local mean velocity averages over stochastic kicks from the medium to quantify how fast those trajectories that pass through angle $x$ at time $t$ tend to be moving.
It can be shown that~\cite{seifertStochasticThermodynamicsFluctuation2012}
\begin{equation}
    \int \dop x \, \nu_t(x) = \frac{\dop \evsm{x}}{\dop t}. \label{eq:nu_to_dxdt}
\end{equation}
In the PSS, the overall mean velocity of the system over a full period is equal to the mean velocity of the control parameter,
\begin{equation}
    v \equiv \ev{\timeavg{\nu_t(x)}} = \xi/\tau,
\end{equation}
where we define the time average of a quantity $A(t)$ as $\timeavg{A} \equiv \tau^{-1} \int \dop t \, A(t)$.

In the PSS, the system entropy change vanishes over any given driving period $\tau$,
\begin{subequations}
\begin{align}
    &\Delta  \Ssys |_\Lambda \equiv \int_t^{t+\tau} \dop t' \, \dot{S}_\mathrm{sys} \\
    &= - \int \dop x \, [p_{t+\tau}^0(x) \ln p_{t+\tau}^0(x) - p_t^0(x) \ln p_t^0(x)] \\
    &= 0 \ ,
\end{align}
\end{subequations}
where $p_t^0(x)$ is the PSS distribution under the particular driving protocol $\Lambda$.
The total entropy production over any cycle is therefore equal to the entropy production in the medium, $\Delta S_\mathrm{tot} |_\Lambda = \Delta S_\mathrm{m} |_{\Lambda}$.
Via~\eqref{eq:ent_heat}, the heat produced over a cycle is therefore
\begin{equation}
    \evsm{Q} =  \, \frac{\tau}{\beta D} \, \ev{\timeavg{\nu^2}}. \label{eq:Q_v_avg}
\end{equation}
In analogy to the identity $\mathbb{E}(x^2) = \mathbb{E}(x)^2 + \Var(x)$ [for generic expectation value $\mathbb{E}(\cdot)$ and variance $\Var(\cdot)$], and via \eqref{eq:nu_to_dxdt}, Eq.~\eqref{eq:Q_v_avg} can be re-written as
\begin{equation}
    \evsm{Q} = \frac{\tau}{\beta D} \left\{ v^2 + \Var_t\left[ \frac{\dop \evsm{x}}{\dop t} \right] + \timeavg{\Var_x[\nu_t(x)]} \right\}. \label{eq:heat_decomposed}
\end{equation}
Here, we define the temporal variance $\Var_t(a) \equiv \timeavg{(a-\timeavg{a})^2}$ and the angular variance $\Var_x(a) \equiv \ev{(a - \evsm{a})^2}$.
We unpack the intuition gained from \eqref{eq:heat_decomposed} about optimal transport and optimal control of periodic systems in Sec.~\ref{sec:theory-oc}.

\subsubsection{Switching Landscapes} \label{sec:st_switching}

In the case of switching landscapes, the work rate is
\begin{equation}
    \evsm{\dot{W}} = \sum_n \int \dop x \, p_t(x,n) \frac{\partial \Ucontrol(x,t)}{\partial t}.
\end{equation}
The system free energy can change via work against the nonequilibrium torque [Eq.~\eqref{eq:free_nrg_single}] or via transitions between landscapes, 
\begin{align}
    \dot{F} = \sum_n  \int \dop x \,  & \bigg[ f \, j_t(x,n) \nonumber \\
    &+ \Delta \mu  \, (R_{n+1,n} - R_{n-1,n}) \, p_t(x,n)\bigg].
\end{align}
Similarly, heat is produced via both the mechanical avenue present in the single-landscape case and an additional avenue from the chemical transitions,
\begin{align}
    \evsm{\dot{Q}} = \sum_n \int \dop x \, \bigg\{ \beta\left[ \frac{\partial \left(\Ucontrol + \Uland^n\right)}{\partial x} - f\right] j_t(x,n&) \nonumber \\ 
    +  Q_\mathrm{chem}^{n+1,n}\, R_{n+1, n}(x)\, p_t(x,n&) \nonumber \\ 
    +  Q_\mathrm{chem}^{n-1,n}  \, R_{n-1, n}(x)\, p_t(x,n&) \bigg\}. \label{eq:heatrate_switching}
\end{align}
We call the first term in curly brackets the mechanical heat $\evsm{\dot{Q}_\mathrm{mech}}$, and the final two terms sum to the chemical heat $\evsm{\dot{Q}_\mathrm{chem}}$.
Due to the chemical heat, \eqref{eq:Q_v_avg} no longer applies to the total heat, but the mechanical heat can still be expressed in terms of a total mechanical entropy production,
\begin{equation}
    \evsm{\dot{Q}_\mathrm{mech}} = \frac{1}{\beta D}\sum_n \int \dop x \, \frac{j_t(x,n)^2}{p_t(x,n)}.
\end{equation}

\subsection{Optimal Control} \label{sec:theory-oc}

We consider the well-known optimal control problem of computing the most efficient schedule for driving a system by manipulating a control parameter between two specified endpoints~\cite{schmiedlOptimalFiniteTimeProcesses2007,sivakEquilibriumMeasurementsNonequilibrium2012,sivakThermodynamicGeometryMinimumdissipation2016,blaberEfficientTwodimensionalControl2022,zhongLimitedcontrolOptimalProtocols2022,guptaOptimalControlF1ATPase2022,warehamMultiparameterOptimalControl2025,chennakesavaluUnifiedGeometricFramework2023,engelOptimalControlNonequilibrium2023,luceroOptimalControlRotary2019,sawchukGlobalThermodynamicManifold2026}.
The cost to be optimized is the mean work done on the system over the course of the protocol.
Formally, the optimization problem is
\begin{subequations}
\begin{align}
    \Lambda^* &\equiv \argmin_{\Lambda} \left.\int_0^\tau \dop t \, \evsm{\dot{W}} \right|_{\Lambda} \label{eq:opt_problem}\\
    &= \argmin_{\Lambda} \int_0^\tau \dop t \int \dop x \, p_t(x) \frac{\partial U(x|\lambda)}{\partial \lambda} \frac{\dop \lambda(t)}{\dop t},
\end{align}
\end{subequations}
for \textit{minimum-work protocol} $\Lambda^*$, where the protocol duration $\tau$ is held fixed.

This optimization problem is difficult because the probability distribution $p_t(x)$ in the integrand depends in general both on the initial condition $p_{t=0}(x)$ and on the entire history of the driving protocol $\lambda(t')$, $t'\leq t$.
To make the problem tractable, simplifying assumptions are often applied, such as an equilibrated initial probability distribution, $p_{t=0}(x)\equiv p_{eq}(x|\lambda(0))$ and/or a separation of timescales between the protocol duration and the system relaxation time~\cite{sivakThermodynamicMetricsOptimal2012,sivakThermodynamicGeometryMinimumdissipation2016,blaberStepsMinimizeDissipation2021,blaberEfficientTwodimensionalControl2022,guptaOptimalControlF1ATPase2022,warehamMultiparameterOptimalControl2025,zhongLinearResponseEquivalence2024,blaberOptimalControlStochastic2023}.
We take a different natural simplifying assumption, namely that we seek optimal protocols for driving a PSS $p_t(x) = p_t^{0}(x)|_{\Lambda}$ [recalling the Floquet-eigenvalue notation for the PSS from \eqref{eq:floquet_limit}].
In principle, $p_t^{0}(x)|_{\Lambda}$ can be calculated directly for each driving protocol by discretizing the probability distribution and operators in the eigenproblem \eqref{eq:eigenproblem}.
Due to technical constraints, we will approximate the PSS by simulating the Fokker-Planck equation until steady state is reached, before computing the work.

All three terms in curly brackets in \eqref{eq:heat_decomposed} are non-negative.
For a single landscape, considering the conditions that minimize individual right-hand side terms of the mechanical heat decomposition~\eqref{eq:heat_decomposed} provides useful intuition about optimal transport and control.
The first [Eq.~\eqref{eq:heat_ot}, below] is fixed given the angular period $\xi$ and the protocol duration $\tau$ (i.e., given the control problem at hand).
We will soon argue that an optimal-transport map reduces the remaining terms to zero, so we call the nonvanishing portion the \textit{optimal-transport heat}~\cite{zhangWorkNeededDrive2020}
\begin{equation}
    \evsm{Q_\mathrm{OT}} \equiv \frac{\tau}{\beta D} v^2 \ , \label{eq:heat_ot}
\end{equation}
with the corresponding uniform-in-time optimal-transport heat rate $\evsm{\dot{Q}_\mathrm{OT}} \equiv v^2/(\beta D)$.~\footnote{Zhang~\cite{zhangWorkNeededDrive2020} defined $\evsm{\dot{Q}_\mathrm{OT}}$ as the ``mechanical'' portion of the dissipated work, $W^*_\mathrm{mech}(t)$, required to drive a system's mean a distance $\xi$ in a time $t$, and called the remaining proportion of the dissipated work the ``thermal'' portion.}

A useful analogy to \eqref{eq:heat_decomposed} is the frictional heat $Q_{\rm det}$ dissipated by a deterministic, low-Reynolds-number object with friction coefficient $\gamma$, that moves through a fluid along a path $x(t)$ with mean velocity $\timeavg{\dop x/\dop t} =v$ over a time $\tau$,
\begin{equation}
    Q_\mathrm{det} = \tau \gamma \left\{v^2 + \Var_t\left[ \frac{\dop x(t)}{\dop t}\right] \right\}. \label{eq:heat_det}
\end{equation}
When taken together, the first two terms in \eqref{eq:heat_decomposed} give the heat predicted by treating the stochastic system as if it moves deterministically along with the ensemble mean.
The temporal-variance term in \eqref{eq:heat_decomposed} can be reduced to zero by ensuring that the mean velocity is uniform in time, $\dop \evsm{x} /\dop t = v$, which is the same result as for a deterministic particle in a viscous fluid.

The third term in curly brackets in \eqref{eq:heat_decomposed} is nonzero for protocols that change the shape of the probability distribution.
Intuitively, such shape changes are associated with an angular dependence on how fast trajectories on average are moving when they cross a point.
To set the integrand $[\nu_t(x) - \evsm{\nu}]^2$ of the time average $\timeavg{\Var_x[\nu_t(x)]}$ to zero at a particular time $t'$, we must have $\nu_{t'}(x) = c_t$, a constant, for all $x$.
From the definition~\eqref{eq:nu_def}, this implies $j_{t'}(x) = c_t' \, p_{t'}(x)$, which implies that at $t'$ the Fokker-Planck equation~\eqref{eq:FPE_flux} reduces to the one-dimensional wave equation 
\begin{equation}
\partial_t \, p_t(x)|_{t'} = - c_t \, \partial_x \, p_t(x)|_{t'}, \label{eq:1D_wave}
\end{equation}
for which solutions are rotating waveforms that do not change shape.
The full contribution of this final term to the heat can be reduced to zero by ensuring that the probability distribution does not change shape throughout the driving protocol, though this does not necessitate that $\nu_t(x)$ is constant in time [i.e., the group velocity $c$ in \eqref{eq:1D_wave} could vary with time].

Taking the two above conditions together, \eqref{eq:heat_decomposed} can be minimized for a fixed $v$ by ensuring that the entire probability distribution slides at a fixed velocity $v$, maintaining its original shape.
A quadratic potential translating with a constant velocity, for example, can achieve both conditions~\cite{mazonkaExactlySolvableModel1999}.
In general, however, these conditions can only be reached with full control over the flux $j_t(x)$; the above argument therefore is most accurately a solution for periodic optimal transport.
When lacking full control, the driving protocol should come as close as possible to achieving these conditions.

There is a deep existing literature on the solution to the optimal-transport problem of minimizing \eqref{eq:Q_v_avg} given full control over the velocity distribution $\nu_t(x)$~\cite{aurellOptimalProtocolsOptimal2011,aurellRefinedSecondLaw2012,seifertEntropyProductionStochastic2005,zhangWorkNeededDrive2020,zhangOptimizationStochasticThermodynamic2020,chennakesavaluUnifiedGeometricFramework2023,zhongLinearResponseEquivalence2024}.
We emphasize that the minimization argument above is specific to the case of identically shaped initial and final distributions~\cite{zhangWorkNeededDrive2020}; the more general case is that the optimal-transport cost is proportional to the $L^2$-Wasserstein metric~\cite{benamouComputationalFluidMechanics2000, aurellOptimalProtocolsOptimal2011,aurellRefinedSecondLaw2012}.
The decomposition we present in \eqref{eq:heat_decomposed} is a useful one, though, and allows heat production to be intuitively connected to the shape of the distribution throughout the driving cycle.

For switching landscapes, there is an additional source of heat production $\evsm{Q_\mathrm{chem}}$ due to the chemical transitions, given by the final two terms in \eqref{eq:heatrate_switching}.
These terms are large when chemical transitions tend to lower the system energy, i.e., when $R_{n\pm 1,n}(x) p_t(x,n)$ is significant in areas with $\Uland^{n\pm1} (x) \pm \Delta \mu > \Uland^n (x)$.
Protocols that reduce heat should seek to limit the proportion of chemical transitions that significantly decrease the system's energy.

\section{Methods} \label{sec:methods}

We developed custom Python code for simulating the Fokker-Planck equations~\eqref{eq:FPE_op} and~\eqref{eq:FPE_switching} with JAX~\cite{jax2018github}, a library for automatic differentiation (AD).
Briefly, this code approximates the functional derivative $\delta\evsm{W}/\delta [\lambda(t)]$ of the work with respect to the control protocol and seeks the solution to the optimization problem \eqref{eq:opt_problem} via the Adam optimizer~\cite{kingmaAdamMethodStochastic2017}.
Since the simulations directly compute $p_t(x)$, the range of scenarios accessible to the method are limited only by the computational cost and/or the stability of the simulations.
We detail the computational methods below; in addition, the code~\cite{warehamGithub2026} and data~\cite{wareham_2026_22261180} is freely available.

The simulations discretize the interval $x \in [0, 2\pi)$ as $N$ points separated by angular separation $\Delta x$, which we denote
\begin{equation}
    x \sim x_j \equiv j \, \Delta x,~j \in \{0,...,N-1\},~\Delta x \equiv 2 \pi/ N. \label{eq:x_disc}
\end{equation}
The periodic boundary condition imposes $x_{k+N} \equiv x_k$, $k \in \mathbb{Z}$, and similar for other discretized, periodic quantities.
Time $t \in [0, \tau)$ is similarly broken into $M$ equally spaced points,
\begin{equation}
    t \sim t^m \equiv m\, \Delta t,~m \in \{0,...,M-1\},~\Delta t \equiv \tau/M.
\end{equation}
We adopt the convention that lower indices refer to angle, while upper indices refer to time.

The probability $p_t(x)$ and potentials are discretized in angle and time as
\begin{subequations}
\begin{align}
  p_t(x) &\sim p_j^m,\\
    \Uland(x) &\sim [\Uland]_j \equiv U(x_j),\\
    \Ucontrol(x,t) &\sim [\Ucontrol]_j^m \equiv \Ucontrol(x_j, t^m).
\end{align}
\end{subequations}
We approximate the periodic, continuous potentials by a truncated Fourier series,
\begin{align}
    \Uland(x) &= \sum_{\kappa = 1}^{N_\mathrm{L}} a^\mathrm{L}_\kappa \cos \frac{2 \pi \kappa}{\xi} x+ b^{\mathrm{L}}_\kappa \sin\frac{2 \pi \kappa}{\xi} x, \label{eq:Uland_fourier} \\
    \Ucontrol(x,t) &= \sum_{\kappa = 1}^{N_\mathrm{C}} a^\mathrm{C}_\kappa \cos \frac{2 \pi \kappa}{\chi} [x - \lambda(t)] \nonumber \\
    &\quad + b^\mathrm{C}_\kappa \sin \frac{2 \pi \kappa}{\chi} [x - \lambda(t)] \ . \label{eq:Ucontrol_fourier}
\end{align}
In general, the control potential's Fourier coefficients ($a^\mathrm{C}_\kappa$ and $b^\mathrm{C}_\kappa$) can also depend periodically on $\lambda(t)$, but here we restrict our attention to control potentials that rotate without changing shape.
Since angular derivatives of the potentials can easily be computed analytically from \eqref{eq:Ucontrol_fourier}, we use those analytical forms at each $x_j$ in the simulations.
Other angular derivatives in the angular Fokker-Planck operator~\eqref{eq:FPE_op} are approximated with (periodic) central differences.

We represent the periodic control protocols with cubic splines between $N_\mathrm{k}$ discrete points (knots) $\{(s, \lambda)\} = \{(s^{(i)}, \lambda^i)\}$~\cite{burdenNumericalAnalysis2015}, for the duration-normalized time $s \equiv t/\tau$.
In each interval $s^{(i)} \leq s < s^{(i+1)}$, this piecewise function has form
\begin{equation}
    \lambda(s) = \xi s + a^i + b^i (s - s^{(i)}) + c^i (s - s^{(i)})^2 + d^i (s - s^{(i)})^3.\label{eq:center_param}
\end{equation}
The parameters $\{(a^i,\, b^i,\, c^i,\, d^i)\}$ are set to ensure continuity of $\lambda(s)$, $\dop \lambda/\dop s$, and $\dop^2 \lambda/\dop s^2$ at all the knots, including across the periodic boundary between $s^{(N_\mathrm{k}-1)}$ and $s^{(0)}$.
The knot-independent linear term is included to ensure that $\lambda(s)$ proceeds ``forward'' by one period over the course of the protocol.
We fix each $s^{(i)}$ at equally spaced times between $s^{(0)} = 0$ and $s^{(N_\mathrm{k}-1)} = 1 - 1/N_\mathrm{k}$.
Furthermore, we have the freedom to set all protocols to pass through $\lambda = 1/2$ rot at $s=1/2$.
The optimization algorithm seeks the choice of the remaining $\{\lambda^i\}$ to minimize the work; in other words, at each iteration our JAX code computes the gradient $\nabla_{\lambda^i}\evsm{W}$ and uses it to update the protocol.
The polynomial coefficients $\{(a^i,\, b^i,\, c^i,\, d^i)\}$ in \eqref{eq:center_param} are fixed given a particular set $\{\lambda^i\}$.

The time-evolution algorithm provides a recipe for computing $p_j^{m+1}$ given $p_j^m$.
We map the angular probability distribution to a column vector $[p]^m$ (writing matrices in square brackets, and with angular index $j$ running over the rows of $[p]^m$), so the Fokker-Planck operator~\eqref{eq:FP_op} for $\lambda^m$ is approximated as a nearly tridiagonal (tridiagonal plus off-diagonal corners) matrix $[\mathcal{L}]^m$ with periodic boundary conditions.
Approximating the time derivative in \eqref{eq:FPE_op} with forward differences, the angular update rule becomes
\begin{equation}
    [p]^{m+1} = \left(I + \Delta t \, [\mathcal{L}]^m\right)[p]^m, \label{eq:step_mech}
\end{equation}
where $I$ is the identity matrix.

For switching landscapes, the chemical transition rates are discretized at each angle via \eqref{eq:detbal}.
We assume that the angular coordinate remains constant during transitions in the chemical coordinate, and adopt a master-equation approach for the time evolution of the landscape state.
This amounts to defining a nearly tridiagonal periodic matrix $[R]_j$ from the rates at each angle, which acts on column vectors with rows that correspond to different landscape states, $[{p_n}]_j^m.$
The rates' exponential dependence on the energy difference between chemical states means that the condition number of $[R]_j$ is often large (i.e., the system of differential equations is stiff), and we must use a backward-difference update scheme, defined by solving the matrix equation
\begin{equation}
    (I - \Delta t [R]_j) \, [p_n]_j^{m+1} = [p_n]_j^{m} \label{eq:step_chem}
\end{equation}
for $[p_n]_j^{m+1}$ at each angular index.
We break the time evolution into a three-step process at each timestep: first, the control protocol takes a discrete step from $[\Ucontrol]^{m-1}$ to $[\Ucontrol]^{m}$ according to the control protocol; next, the probability distribution relaxes along the angular coordinate~[Eq.~\eqref{eq:step_mech}]; finally, the probability distribution relaxes along the chemical coordinate~[Eq.~\eqref{eq:step_chem}].

We define the work done by the potential over a timestep as the change in average energy induced by incrementing the control potential, with the system held fixed:
\begin{multline}
    \int_{t_{m-1}}^{t_m} \dop t' \, \evsm{\dot{W}(t')} \approx \ev{W}^m \\
    \equiv  \sum_n \Delta x\sum_j \{[\Ucontrol]_j^{m} - [\Ucontrol]_j^{m-1}\} (p_n)_j^{m-1}. \label{eq:work_inc}
\end{multline}
The sum over angular index is equivalent to trapezoid-rule numerical integration for a periodic domain.

Similarly, we define the mechanical heat increment as the change in energy due to the angular relaxation step,
\begin{multline}
    \int_{t_{m-1}}^{t_m} \dop t' \, \evsm{\dot{Q}_\mathrm{mech}(t')} \approx \evsm{Q_\mathrm{mech}}^m \\
    \equiv \, \sum_n \Delta x \sum_j [U^n]_j^m \left\{[p_n]_j^{m-1/2} - [p_n]_j^{m-1} \right\},
\end{multline}
where the half-time-indexed probability $[p_n]_j^{m-1/2}$ denotes the probability after the angular update and before the chemical update.

The chemical heat increment is the change in energy due to the chemical-reaction step.
We use an implicit update rule for this step, so in principle computing the change in probability over the chemical-reaction step $[p_n]_j^{m} - [p_n]_j^{m-1/2}$ leaves some ambiguity about where the probability density shifted (to/from either $n-1$ or $n+1$), and therefore some ambiguity about the change in energy each of those transitions incurred.
In practice, though, we consider cases for which transition events occur where there is a large difference in the fluxes through these transitions at each given point (e.g., $n\to n+1$ is highly probable, while $n+2\to n+1$ is very rare), so we can assign that change in probability density at a point to the single most probable transition.
The chemical heat produced over the course of the protocol is
\begin{multline}
    \int_{t_{m-1}}^{t_m} \dop t' \, \ev{\dot{Q}_\mathrm{chem}(t')} \approx \evsm{Q_\mathrm{chem}}^m \\
    \equiv \sum_n \Delta x \sum_j [\Theta^n]_j \, \left\{[\Uland^n]_j - [\Uland^{n+1}]_j - \Delta \mu \right\} \\
    \times \left\{ [p_n]_j^{m} - [p_n]_j^{m-1/2} \right\}.
\end{multline}
The transition window indicator $[\Theta^n]_j$ is either zero or one, and picks out the angular region in which transitions between state $n$ and $n+1$ occur with non-negligible probability.

Finally, the system entropy increment is
\begin{multline}
    \int_{t_{m-1}}^{t_m} \dop t' \, \ev{\dot{S}_\mathrm{sys}(t')} \approx \ev{S_\mathrm{sys}}^m \\
    \equiv \sum_n \Delta x \sum_j \left( [p_n]_j^{m} \ln [p_n]_j^{m} - [p_n]_j^{m-1} \ln [p_n]_j^{m-1} \right) \ .
\end{multline}
The average work, heat, and entropy change over the course of a protocol are computed by summing the respective increments over the time indices $m$.

In Appendix~\ref{app:langevin}, we compare input work $\evsm{W_\mathrm{in}}$ computed with our Fokker-Planck simulations to that computed by averaging an ensemble of Langevin simulations, when driven under our naive, linear-response-designed, and AD-designed optimal protocols.
In all cases, we find strong agreement between the two methods.

The code computes the cost $\ev{W}$ via \eqref{eq:work_inc} for a system in the PSS corresponding to $\Lambda$; the automatically differentiated code generated by JAX computes $\delta\ev{W}/\delta [\lambda(t)]$.
After initializing with a Gaussian probability distribution centered in the well of the trapping potential, the PSS is computed by simulating the system for sufficiently many full $2 \pi$ rotations of the control parameter to qualitatively reproduce the same probability distribution and thermodynamic flows throughout further driving periods.
The control protocol is initialized as the naive protocol ($\lambda^i = 0$) and iteratively updated via the Adam optimizer~\cite{kingmaAdamMethodStochastic2017} toward the optimal protocol.

We compare our results to the near-equilibrium optimal protocol computed via linear-response theory~\cite{sivakThermodynamicMetricsOptimal2012,zulkowskiOptimalControlOverdamped2015,blaberOptimalControlStochastic2023}.
The linear-response optimal protocol (which does not depend on protocol duration) is designed by computing the friction tensor $\zeta(\lambda)$, which acts as a metric on the control manifold.
In the case of a single control parameter, the optimal protocol is specified by enforcing the proportionality $\dop \lambda^* / \dop t \propto \zeta(\lambda)^{-1/2}$, with normalization set to ensure that the control parameter completes exactly one cycle over the protocol duration $\tau$.
We compute $\zeta(\lambda)$ according to the numerical procedure outlined in Refs.~\cite{guptaOptimalControlF1ATPase2022,warehamMultiparameterOptimalControl2025}, which involves extracting equilibrium properties of the system from long Langevin simulations with $\lambda$ held fixed.

\section{Results} \label{sec:results}

In this section, we present efficient protocols designed for simple models of each of the single- and switching-landscape cases, and analyze the design principles that these protocols exhibit.
These simple models are ideal for seeking broad design principles for efficient control of periodic Brownian systems.

\subsection{Single Landscape} \label{sec:results-single}

\begin{figure}
    \centering
    \includegraphics[width=\columnwidth]{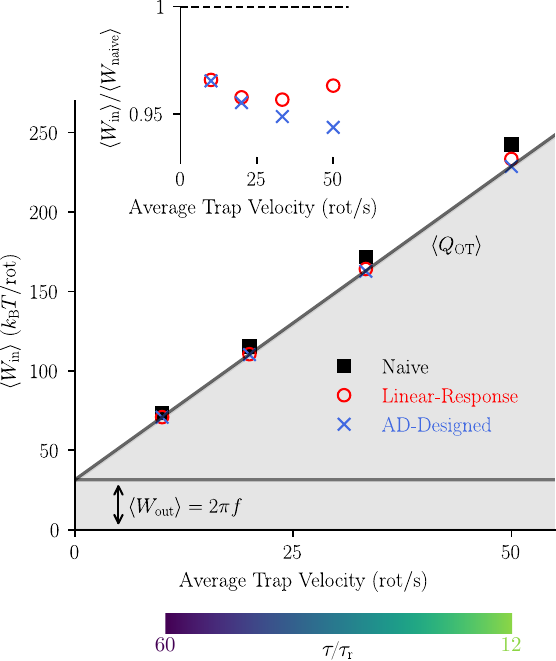}
    \caption{Average input work $\ev{W_\mathrm{in}}$ required to drive one rotation for the single landscape [Sec.~\ref{sec:results-single}] under the naive protocol (black squares), linear-response protocol (red circles), and automatic-differentiation-designed protocols (blue crosses). The AD-designed protocols approach the optimal-transport limit $\evsm{W_\mathrm{out}} + \evsm{Q_\mathrm{OT}}$ (black diagonal line) for all driving speeds we consider here. 
    Horizontal color bar below $x$-axis indicates corresponding protocol duration $\tau/\tau_{\rm r}$. 
    Inset: Linear-response and AD-designed protocol works, normalized at each driving speed to the naive work. At slow driving speeds, linear response is nearly as efficient as the AD-designed protocol. However, for faster driving which drives further out of equilibrium, the AD-designed protocols outperform linear response; here, by up to $\sim$2$\%$.}
    \label{fig:single-opt}
\end{figure}

\begin{figure}
    \centering
    \includegraphics[width=\columnwidth]{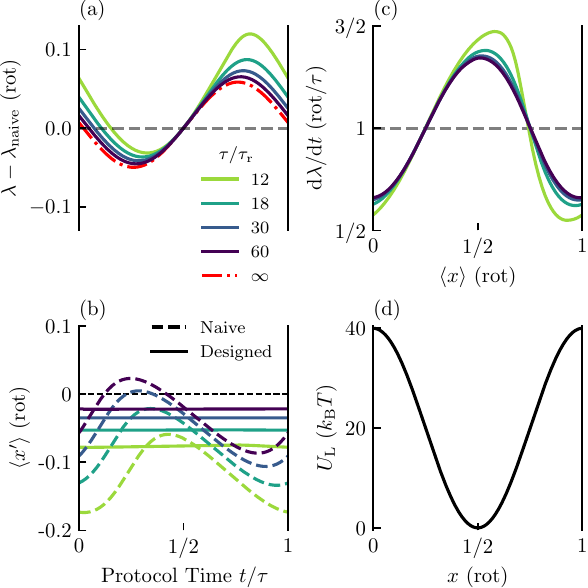}
        \caption{Designed protocols for the single landscape at various protocol durations $\tau$, presented in units of the control-potential relaxation time $\tau_\mathrm{r}$. All protocols pass through $\lambda = 1/2$ rot at $t/\tau = 1/2$. (a) Deviation of the designed protocols (solid curves) from the naive, constant-velocity protocol $\Lambda_\mathrm{naive}$ (horizontal dashed gray line). The slowest protocols approach the linear-response prediction (red dash-dotted curve), while faster protocols are increasingly skewed (time-reversal asymmetric). (b) Mean bead angle in the comoving reference frame $\ev{x'}$ (i.e., with the naive trap angle defining $x'=0$), for naive (dashed curves) and designed (solid curves) protocols. The designed protocols rotate the bead mean at a near-constant velocity. (c) Parametric plot of designed-protocol control-parameter velocity $\dop \lambda/\dop t$ against mean bead angle $\ev{x}$. (d) Schematic of the landscape potential $\Uland$. Relative to the naive protocol [horizontal dashed gray line in (c)], designed protocols slow down when the bead is in the metastable barrier region.}
    \label{fig:single-prots}
\end{figure}

First, we examine the case of a single landscape potential with no chemical transitions.
We consider the lowest non-trivial order for the landscape and control potentials,
\begin{subequations}
    \begin{align}
    \Ucontrol &= -A_\mathrm{C} \cos [x - \lambda(t)], \\
    \Uland &= A_\mathrm{L} \cos(x) \ ,
\end{align}
\end{subequations}
with $A_\mathrm{C} = 60~\kB T$, $A_\mathrm{L} = 20~\kB T$, nonequilibrium force $f = 5~\kB T/\mathrm{rad}$, and bead diffusion coefficient $D = 10~\mathrm{rad}^2/s$.
These parameters are of a similar scale to those accessible in experiments on $\mathrm{F_1}$-ATPase~\cite{guptaOptimalControlF1ATPase2022,mishimaEfficientlyDriving2025}.

Figure~\ref{fig:single-opt} shows the average input work $\ev{W_\mathrm{in}}$ required to drive each $2\pi$ rotation of the bead in the periodic steady state, as a function of average trap velocity.
The designed protocols computed with our automatically differentiated Fokker-Planck simulation consistently outperform both the naive protocol and the protocol designed using linear-response theory.
Throughout this range of driving speeds, the AD-designed protocols reduce the input work to the limit $\evsm{W_\mathrm{out}} + \evsm{Q_\mathrm{OT}}$ predicted by optimal transport, shown in the black line.
As expected, the linear-response-designed protocol performs similarly to AD-designed for the slowest protocol we consider, and our AD-designed protocols outperform linear response by an increasing margin for faster driving.

Figure~\ref{fig:single-prots}(a) shows the time of course of the designed protocols and the linear-response protocols. 
To emphasize their differences, we present the protocols as variations around the constant-velocity naive protocol; i.e., we show the protocols in the comoving frame of reference defined in Sec.~\ref{sec:theory-models}.
The designed protocols are labeled with the protocol duration in units of the relaxation time $\tau_\mathrm{r} = 1/(A_\mathrm{C} D) = (1/600)~\mathrm{s}$ for a bead with diffusion coefficient $D$ in a quadratic potential with stiffness $A_\mathrm{C}$ (the first-order Taylor-series approximation to $\Ucontrol$).
Unlike the system relaxation time under the combined potential $\Ucontrol + \Uland$, this quadratic relaxation time does not vary with $\lambda$, so it provides a representative static timescale at all control-parameter values.

As a useful test of the validity of our method, our AD-designed protocols approach the linear-response prediction for long protocol durations, despite the optimization algorithm having no explicit knowledge of linear-response theory, and only seeking to minimize $\ev{W_\mathrm{in}}$.
The AD-designed protocols become progressively more time-reversal asymmetric as duration decreases.
This growing deviation from the linear-response prediction accounts for the increasingly far-from-equilibrium driving.

\begin{figure}
    \centering
    \includegraphics[width=\columnwidth]{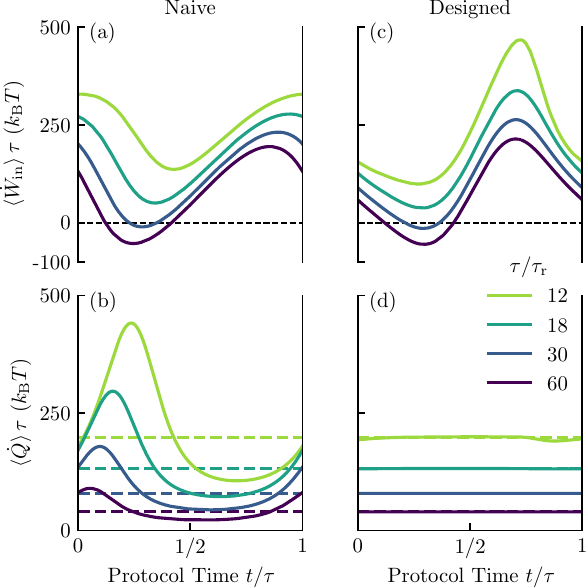}
    \caption{Thermodynamic flows for a single landscape as a function of protocol time. (a) Work and (b) total heat rates for the naive protocol. (c) Work and (d) total heat rates for the designed protocols. The designed protocols seek to set the total heat rate equal to the rate $\evsm{\dot{Q}_\mathrm{OT}}$ predicted under optimal transport [Eq.~\eqref{eq:heat_ot}], shown by the dashed lines in (d).
    All rates are multiplied by the protocol duration $\tau$ so that the integral of each curve (over the dimensionless protocol time on the $x$-axis) yields the cumulative work or heat over the protocol.
}
    \label{fig:single-work-heat}
\end{figure}

Figure~\ref{fig:single-prots}(b) shows the mean bead angle in the comoving frame $\ev{x'}$ as a function of protocol time, for beads driven by the naive protocol and designed protocols.
For both protocol types, as driving speed increases, the bead lags farther behind the trap center located at the origin of the comoving frame~\cite{mazonkaExactlySolvableModel1999}.
All designed protocols maintain a near-constant lag for the bead: $\dop \evsm{x} /\dop t \approx v$.
This minimizes the time-variance term in \eqref{eq:heat_decomposed} for the mechanical heat, as argued in Sec.~\ref{sec:theory-oc}.
We emphasize that this result arises naturally in the minimization of $\evsm{W_\mathrm{in}}$; the minimization procedure has no explicit knowledge of the mechanical heat decomposition~\eqref{eq:heat_decomposed}.

Figure~\ref{fig:single-prots}(c) is a parametric plot of trap velocity $\dop \lambda/\dop t$ (normalized to the naive-protocol velocity) against mean bead angle.
Comparing with the schematic of the landscape potential shown in Fig.~\ref{fig:single-prots}(d), the trap slows down when the bead's probability distribution is in the ``barrier region'', near the peak of $\Uland$.
This extends beyond linear response the intuition from previous optimal-control studies that control should slow down when the bead is in the barrier region~\cite{sivakThermodynamicGeometryMinimumdissipation2016,guptaOptimalControlF1ATPase2022,blaberOptimalControlStochastic2023,engelOptimalControlNonequilibrium2023,warehamMultiparameterOptimalControl2025}.

Figure~\ref{fig:single-work-heat} presents the thermodynamic flows $\ev{W_\mathrm{in}}$ and $\evsm{Q}$ when driving with naive and designed protocols.
Qualitatively, the work rates appear similar, with the heat rates showing starker differences.
The designed protocols all seek to set the heat rate equal to $\evsm{\dot{Q}_\mathrm{OT}}$ throughout the entire protocol.
In Sec.~\ref{sec:theory-oc}, we argued that this was the optimal condition, and here we reiterate that the only optimization objective is the minimization of $\ev{W_\mathrm{in}}$, so the algorithm learns this design principle on its own.

\subsection{Switching Landscape} \label{sec:results-switching}

We also make a simple choice of potential for our model of a molecular machine with explicit chemical transitions,
\begin{align}
    \Ucontrol &= - A_\mathrm{C} \cos [x - \lambda(t)], \\
    \Uland^n &= - A_\mathrm{L} \cos\left(x - \frac{2 \pi}{3} n\right),
\end{align}
with $A_\mathrm{C} = 80~\kB T$, $A_\mathrm{L} = 30~\kB T$, $n \in \{0,1,2\}$, vanishing nonequilibrium force $f = 0$, and bead diffusion coefficient $D = 10~\mathrm{rad}^2/s$ [producing approximate system relaxation time $\tau_r = (1/800)~\mathrm{s}$ in the control potential].
Chemical transition rates obey Eqs.~\eqref{eq:switching_rates} with $\Gamma = 1000~\mathrm{s}^{-1}$ and $q=0.5$.
The control protocols are $1/3$-rotation periodic, so in Figures throughout we show only one of these periodic images.

\begin{figure}
    \centering
    \includegraphics[width=\columnwidth]{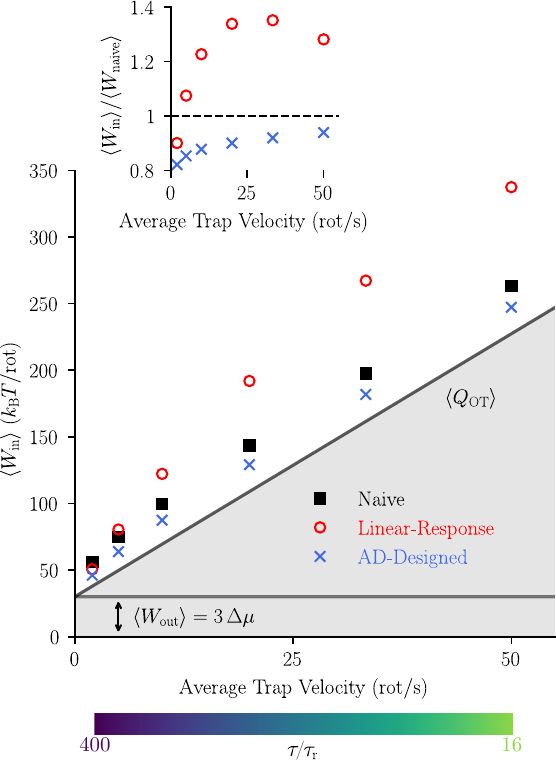}
    \caption{Input work $\ev{W_\mathrm{in}}$ for switching landscapes under the naive protocol (black squares), linear-response protocol (red circles), and automatic-differentiation-designed protocols (blue crosses), as a function of average trap velocity. 
    Horizontal color bar below $x$-axis indicates corresponding protocol duration $\tau/\tau{\rm r}$.
    Inset: Input work for the two designed-protocol types, normalized to the input work $\evsm{W_\mathrm{naive}}$ for the naive protocol at each driving speed. The AD-designed protocols outperform the naive protocol at all driving speeds we consider, whereas the linear-response protocol is ill-suited, only outperforming even the naive protocol in the slowest driving condition. The optimal-transport limit $\evsm{W_\mathrm{out}} + \evsm{Q_\mathrm{OT}}$ (black diagonal line) only accounts for mechanical heat. (Figure~\ref{fig:switching-heat-breakdown} shows that the additional heat in the naive and designed protocols is made up of a combination of both chemical and mechanical heat.)}
    \label{fig:switching-opt}
\end{figure}

Figure~\ref{fig:switching-opt} compares the average input work $\evsm{W_\mathrm{in}}$ required to drive one full rotation under the naive, linear-response-designed, and automatic-differentiation-designed protocols.
For all driving speeds we consider here, the AD-designed protocol significantly outperforms the naive protocol.
Meanwhile, the linear-response-designed protocol is less efficient than naive for all but the slowest driving speed. 
This is not unexpected, as the switching dynamics slows relaxation, hence pushing the system further from equilibrium at a given driving speed. 
The AD-designed protocols' success where linear-response fails highlights the advantage of an optimization framework not reliant on near-equilibrium approximations.

\begin{figure}
    \centering
    \includegraphics[width=\columnwidth]{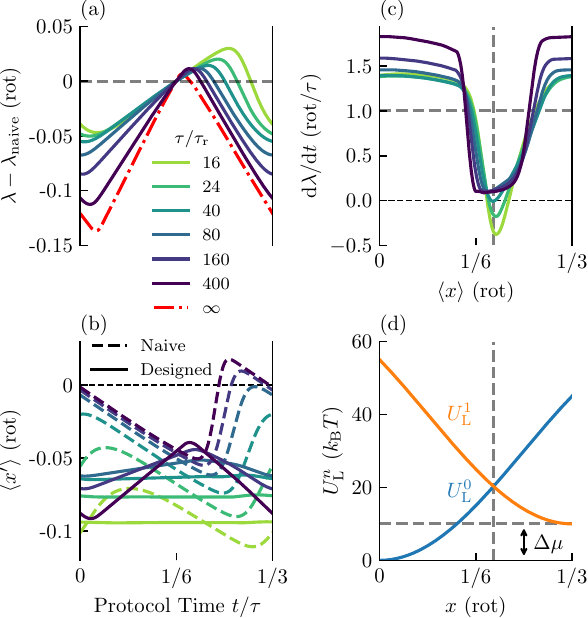}
    \caption{Designed protocols for switching landscapes [Sec.~\ref{sec:results-switching}] at various protocol durations $\tau$, with time in units of the control-potential relaxation time $\tau_\mathrm{r}$.  All protocols pass through $\lambda = 1/6$ rot at $t/\tau = 1/6$. (a) Deviation of the designed protocol (solid curves) from the naive, constant-velocity protocol $\Lambda_\mathrm{naive}$ (dashed gray line). (b) Mean bead angle $\ev{x'}$ in the comoving frame of reference (i.e., with the naive trap angle defining $x'=0$), for naive (dashed curves) and designed (solid curves) protocols with different durations. The designed protocols rotate the bead mean at a near-constant rate, though there is more variation here than for the single landscape [Fig.~\ref{fig:single-prots}(b)]. (c)  Parametric plot of designed-protocol control-parameter velocity $\dop \lambda/\dop t$ against mean bead angle $\ev{x}$. Horizontal dashed gray line: naive protocol. (d) Schematic of the first two landscape potentials $\Uland^n$, including the chemical free-energy change $\Delta \mu$ between adjacent landscapes (horizontal gray line). Relative to the naive protocol, the designed protocols slow down when the bead is in the metastable region near the intersection of neighboring landscapes (vertical dashed gray line in c,d).}
    \label{fig:switching-prots}
\end{figure}

Figure~\ref{fig:switching-prots}(a) shows the deviation of the designed protocols from the naive protocol as a function of protocol time.
As protocol duration $\tau$ increases, the AD-designed protocols progressively bear a stronger resemblance to the linear-response-designed protocol, though the slowest AD-designed protocol for switching landscapes is more distinct from linear response than the slowest AD-designed protocol for the single landscape [Fig.~\ref{fig:single-prots}(a)].
Combined with the protocol performance in Fig.~\ref{fig:switching-opt}, Fig.~\ref{fig:switching-prots}(a) indicates that our learning algorithm for switching landscapes is, in principle, able to design efficient driving protocols in both the near- and far-from-equilibrium regimes.

\begin{figure}[!ht]
    \centering
    \includegraphics[width=\columnwidth]{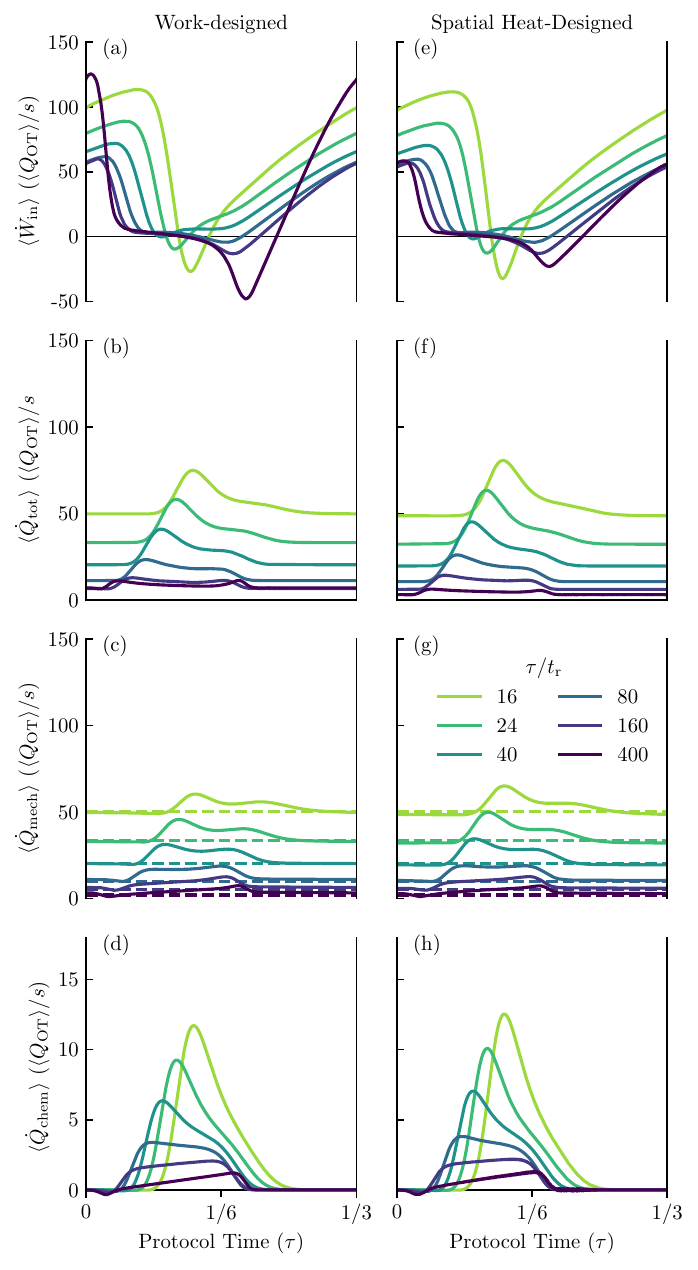}
    \caption{Thermodynamic rates for switching landscapes, multiplied by the protocol duration $\tau$, as a function of protocol time. (a,e) Work flow and (b,f) total heat flow for the (a,b) naive protocol and (e,f) designed protocols. The designed protocols all reduce the total heat flow toward the optimal-transport heat flow $\evsm{\dot{Q}_\mathrm{OT}} \, \tau$ [Eq.~\eqref{eq:heat_ot}, horizontal dashed lines in (g)], but are less successful in this pursuit than for the single landscape [Fig.~\ref{fig:single-work-heat}(d)]. (c,g) Mechanical heat flows. (d,h) Chemical heat flows. The mechanical heat is the dominant contribution to the total heat [note the shorter $y$-axis in (d,h) compared to the other panels].}
    \label{fig:switching-work-heat}
\end{figure}

Figure~\ref{fig:switching-prots}(b) shows the average bead angle $\ev{x}$ relative to the origin of the comoving frame of reference, when driven by the naive or the AD-designed protocol.
The designed protocol is more successful at confining $\dop \evsm{x}/\dop t$ to a small range (most of the designed protocols have smaller variation in the average bead lag than their naive counterparts), as predicted in Sec.~\ref{sec:theory-oc}, though here the variation is larger than for the single landscape [Fig.~\ref{fig:single-prots}(b)].
For slower driving, the mean bead angle more closely tracks the trap-center angle [Fig.~\ref{fig:switching-prots}(a)] reflecting that slower driving allows the bead more time to relax toward its equilibrium position in the driving potential.
The greater variation of duration-normalized velocity $\dop \ev{x}/\dop s$ for the slowly driven protocols, shown in Fig.~\ref{fig:switching-prots}(b), does not lead to a large absolute heat cost, though: that contribution to $\evsm{Q_\mathrm{mech}}$ [from Eq.~\eqref{eq:heat_decomposed}] can be written as $\tau \, \Var (\dop \evsm{x}/\dop t) = \tau^{-1} \, \Var( \dop \evsm{x}/\dop s)$, so duration-normalized velocities scale with the inverse of protocol duration $\tau$, and relative variation in the velocity becomes less costly for slower protocols.

Figure~\ref{fig:switching-prots}(c) is a parametric plot of trap-center velocity $\dop \lambda /\dop t$ of the AD-designed protocols versus the resulting mean bead angle $\evsm{x}$ when driven by those protocols.
Comparing with the landscape potentials plotted in Fig.~\ref{fig:switching-prots}(d), the velocity of all the protocols tends to decrease when the bead is near the intersection of the neighboring landscapes (indicated by the vertical dashed line).
This result is consistent with previous linear-response studies which found that the control-parameter friction coefficient is largest in these transition regions~\cite{guptaOptimalControlF1ATPase2022,warehamMultiparameterOptimalControl2025}.
However, in a major deviation from the prediction of linear-response theory, our fastest AD-designed protocols have a negative trap-center velocity near the intersection.

Figure~\ref{fig:switching-work-heat} compares the average work and heat flows incurred by driving this switching system under the naive (a-d) or AD-designed (e-h) protocols.
Figures~\ref{fig:switching-work-heat}(c,g) show the average heat flow due to the mechanical relaxation of the bead.
Throughout about half of the protocol, the AD-designed protocols achieve the design principle of reducing the mechanical heat rate to the optimal-transport rate $\evsm{Q_\mathrm{OT}}$, but do not achieve this throughout the remaining half of the protocol.
This result indicates that the potential structure considered here is unable to enact sufficient control over the angular probability distribution, and/or that the protocols are designed to balance a tradeoff between the mechanical and chemical heats.
Figures~\ref{fig:switching-work-heat}(d,h) show the heat flow due to chemical transitions.
For all protocol durations, the chemical heat produced by the naive protocol is sharply peaked in time, whereas the designed protocols distribute the chemical heat over a longer interval.
There is a rough correspondence in the times in Fig.~\ref{fig:switching-work-heat}(g) when the designed protocols induce mechanical heat beyond the optimal-transport prediction and the times in Fig.~\ref{fig:switching-work-heat}(h) when chemical heat is produced (indicating that net chemical transitions between energy landscapes with differing energy are occurring).
This correspondence hints that the greatest difficulty in achieving the optimal-transport heat rate occurs when chemical transitions are taking place.
The relative scale of $\evsm{\dot{Q}_\mathrm{mech}}$ and $\evsm{\dot{Q}_\mathrm{chem}}$ indicates that mechanical heat is the dominant avenue for dissipated energy; however, this heat is dominated by the mandatory optimal-transport heat $\evsm{\dot{Q}_\mathrm{OT}}$, especially for faster protocols.

\begin{figure}
    \centering
    \includegraphics[width=\columnwidth]{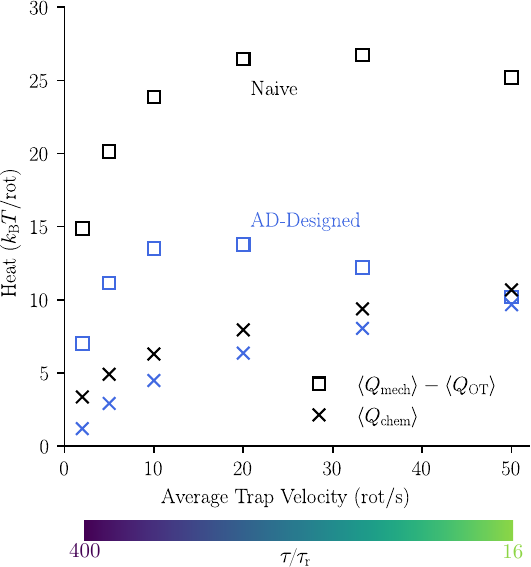}
    \caption{Heat produced for switching landscapes via mechanical relaxation (squares) and via chemical relaxation (crosses), when driven by a naive (black) or designed (blue) protocol. 
    (We subtract the optimal-transport heat $\evsm{Q_{\mathrm{OT}}}$ from the mechanical heat $\evsm{Q_{\mathrm{mech}}}$ to place it on the same scale as the chemical heat $\evsm{Q_{\mathrm{chem}}}$.) 
    The designed protocols reduce the heat produced via either avenue.}
    \label{fig:switching-heat-breakdown}
\end{figure}

Figure~\ref{fig:switching-heat-breakdown} compares the chemical heat $\evsm{Q_\mathrm{chem}}$ to the additional mechanical heat beyond that of the optimal-transport limit, $\evsm{Q_\mathrm{mech}} - \evsm{Q_\mathrm{OT}}$.
These quantities are of a comparable scale for both the naive and designed protocols.
Across all examined protocol durations, the designed protocol reduces the heat through each channel.
At all but the fastest driving speed, both the absolute value and the relative savings in the mechanical heat are larger than the chemical heat.
Appendix~\ref{app:mech-minimizer} shows that protocols designed to minimize the mechanical heat $\evsm{Q_\mathrm{mech}}$ also reduce the chemical heat $\evsm{Q_\mathrm{chem}}$ as a byproduct, and that any additional savings in $\evsm{Q_\mathrm{mech}}$ that are produced by optimizing only this quantity are small.

\begin{figure}
    \centering
    \includegraphics[width=\columnwidth]{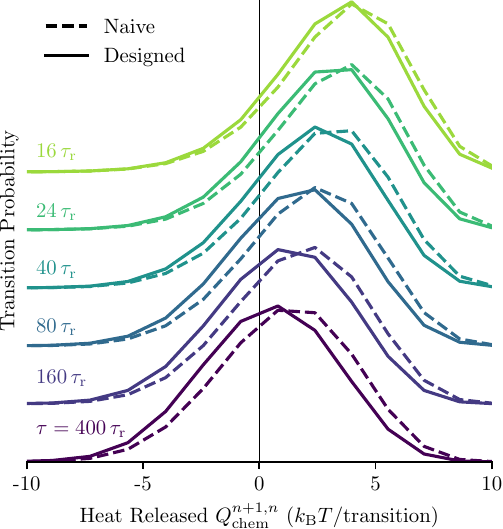}
    \caption{Probability distributions of heat released during chemical transitions, for different protocol durations $\tau$. Curves are offset for greater visual clarity. Faster driving leads to chemical transitions at greater cost (shifting the distribution to the right), since faster rates are required to achieve the same number of transitions in the shorter protocol duration [Eqs.~\eqref{eq:switching_rates}]. In all cases, the designed protocol shifts the distributions to the left, reducing average chemical heat.}
    \label{fig:switching-energies}
\end{figure}

Figure~\ref{fig:switching-energies} shows the probability distribution of heat $Q_\mathrm{chem}^{n+1,n}$ released during chemical transitions, for varying durations and protocol types.
As driving speed increases, transitions tend to be more costly (occurring in regions with a larger heat release for synthesis, and therefore with chemical rates more biased toward synthesis) to achieve the same number of transitions in the shorter driving period.
For all driving speeds, the designed protocols shift the distribution to less positive heat release, amounting to an overall reduction in mean heat.
These results are consistent with our reasoning in Sec.~\ref{sec:theory-oc} regarding reduction of the chemical heat.
They also provide additional motivation for the trap center to slow down when the angular probability distribution is near a transition energy of zero, i.e., at the intersection between potentials [as shown in Figs.~\ref{fig:switching-prots}(c,d)]: Holding the probability distribution for as long as possible near the potential intersection allows more opportunity for less costly chemical transitions to take place.

\section{Discussion} \label{sec:discussion}

In this paper, we presented energy-efficient control protocols for simple models of molecular machines, 
designed using automatic differentiation of Fokker-Planck dynamics. 
Such designed protocols display intuitive design principles to be tested in more realistic systems.
The existing code is easily applied to more accurate models of molecular machines, like the TASAM model of an $\mathrm{F_1}$-ATP synthase driving experiment derived in Ref.~\cite{kawaguchiNonequilibriumDissipationfreeTransport2014} and for which linear-response-designed protocols were computed in Refs.~\cite{guptaOptimalControlF1ATPase2022,warehamMultiparameterOptimalControl2025}.
With some small modifications, the same codebase could account for more complicated chemical rates [a different choice for Eqs.~\eqref{eq:switching_rates}] or multiple control parameters.

In this paper, we argue that efficient control protocols between endpoint distributions with the same shape---such as a periodic steady state---should seek to achieve a constant mechanical heat rate equal to the optimal-transport heat rate, $\evsm{\dot{Q}_\mathrm{mech}} \approx \evsm{\dot{Q}_\mathrm{OT}}$.
We derived an intuitive connection of this stochastic result to efficient transport of deterministic overdamped particles, which dissipate minimal heat when transported at a constant speed.
This design principle is a special case of the result that optimal-transport maps linearly transport all quantiles of the distribution between its endpoints~\cite{aurellOptimalProtocolsOptimal2011,aurellRefinedSecondLaw2012,zhangWorkNeededDrive2020,zhangOptimizationStochasticThermodynamic2020,chennakesavaluUnifiedGeometricFramework2023}.

The designed protocols for the single landscape effectively achieve the optimal-transport heat limit across all considered driving speeds.
This could be due to the particularly simple control problem we consider (with both the control and landscape potential having a sinusoidal shape).
The designed protocols for switching landscapes are significantly worse at controlling the angular distribution to achieve the optimal-transport limit for the mechanical heat, struggling in particular with controlling the shape of the distribution when the system is undergoing transitions between chemical states.
We hypothesize that the simple structure of this control potential is insufficient to maintain the shape of the distribution in the face of the opposing forces experienced by the coexisting populations in different chemical states.
Appendix~\ref{app:mech-minimizer} shows that the AD-designed protocols achieve mechanical heat production close to the minimum achieved when optimizing for only $\evsm{Q_\mathrm{mech}}$, indicating that the additional heat in $\evsm{Q_\mathrm{mech}}$ exceeding the optimal-transport limit does not stem from a tradeoff between reducing the mechanical heat $\evsm{Q_\mathrm{mech}}$ and the chemical heat $\evsm{Q_\mathrm{chem}}$ in order to reduce their sum.

A less precise, but more immediately actionable design principle illustrated by our results is that trap-position control should slow down when the system's probability distribution occupies a metastable ``barrier region,'' near the peak of a continuous potential or where transitions between chemical states tend to occur.
Linear-response theory yields the similar intuition that near-equilibrium control protocols should slow down when the equilibrium distribution (given a particular trap position) is in this barrier region.
Our result is essentially the same concept, but generalized beyond the assumption that the driven probability distribution is well-approximated by equilibrium.

Models that include chemical transitions between potentials have an additional avenue for heat production, via the change in energy associated with the transition.
The AD-designed protocols reduce the heat produced via this avenue by shifting the distribution of chemical transitions toward near-zero energy change.

In this paper, we limited our attention to cases where trajectories that escape the local trap are vanishingly rare, so that all driving protocols achieve the same output work $\evsm{W_\mathrm{out}}$.
In these cases, the input work $\evsm{W_\mathrm{in}} = \evsm{W_\mathrm{out}} + \evsm{Q}$ is a useful measure of the protocols' relative performance.
However, this cost function does not directly measure the quantity of output work (e.g., the proportion of trajectories which actually make a chemical transition).
If there is a significant population of trajectories that escape the local trap (therefore incurring much lower input work), the learning algorithm could learn to exploit that `slip' and design protocols that do minimal input work, but also very little useful output work.
Pushing this optimization procedure to cases where slip occurs is an interesting extension.
With an appropriately generalized cost function~\cite{whitelamHowTrainYour2023}, this method could be effective for designing optimal control in situations where ineffective cycles are present.

The computational cost of designing an optimal protocol using our code scales with the fineness of the angular discretization and with the number of simulated timesteps.
Shorter-duration protocols therefore require less resources to design (assuming the same angular-temporal grid spacing).
Linear-response theory is most effective for designing longer-duration, near-equilibrium protocols, making the two methods complementary.

Designing control protocols for out-of-equilibrium systems is analytically challenging (in all but the most simple cases) because the system's probability distribution depends non-trivially on the control history.
Even for periodic control, analytically solving the Floquet problem~\eqref{eq:eigenproblem} for the periodic steady state is challenging.
An automatically differentiated simulator like the one we present here accounts for this issue by implicitly optimizing over the periodic steady states that the control protocol produces.
No \textit{a priori} knowledge of the `correct' initialization state is required; this optimization is taken care of by the same machinery that designs the rest of the protocol.
Codes like this one may be useful in other instances where some optimization over the system's initial state is possible and desired.

\begin{acknowledgments}
We thank Antonio Patr\'on-Castro and Jordan Sawchuk (SFU Physics), and Shoichi Toyabe, Yohei Nakayama, and Takahide Mishima (Tohoku University) for helpful discussions that improved this paper.
This work was supported by the Natural Sciences and Engineering Research Council of Canada (NSERC), via the following grants: CGS Master's and Doctoral Fellowships and a Michael Smith Foreign Study Supplement (W.C.W.); and an Alliance International Collaboration Grant ALLRP-2023-585940, a Discovery Grant and Discovery Accelerator Supplement RGPIN-2020-04950, and a Tier-II Canada Research Chair CRC-2020-00098 (D.A.S.).
Computing resources were provided by the BC DRI Group and the Digital Research Alliance of Canada (www.alliancecan.ca).
\end{acknowledgments}

\textbf{Data Availability Statement} --- Code~\cite{warehamGithub2026} and data~\cite{wareham_2026_22261180} for this paper is openly available.

\appendix

\begin{figure}[h]
    \centering
    \includegraphics[width=\columnwidth]{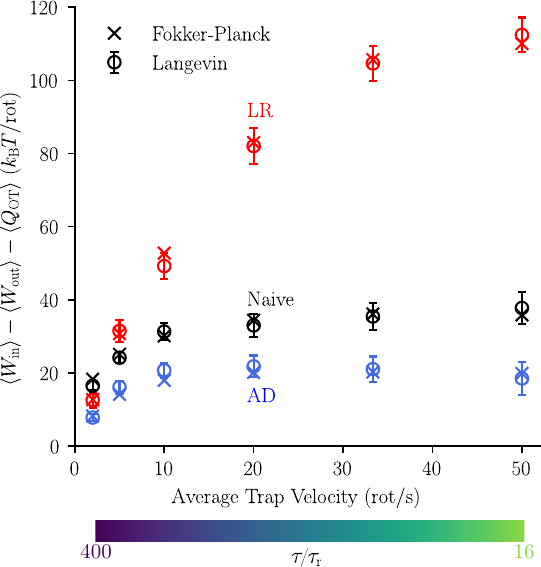}
    \caption{Wasted energy $\evsm{W_\mathrm{in}} - \evsm{W_\mathrm{out}} - \evsm{Q_\mathrm{OT}}$ from Fokker-Planck (crosses) and Langevin (circles) simulations, under naive (black), LR-designed (red) and AD-designed (blue) protocols. Error bars on the Langevin simulations show two standard errors of the mean.
    We plot the wasted energy to remove the dominant trend of the minimum work $\evsm{W_{\mathrm{out}}}+\evsm{Q_{\mathrm{OT}}}$ required to drive a rotation at the given speed.}
    \label{fig:langevin-compare}
\end{figure}

\begin{figure}[t]
    \centering
    \includegraphics[width=\columnwidth]{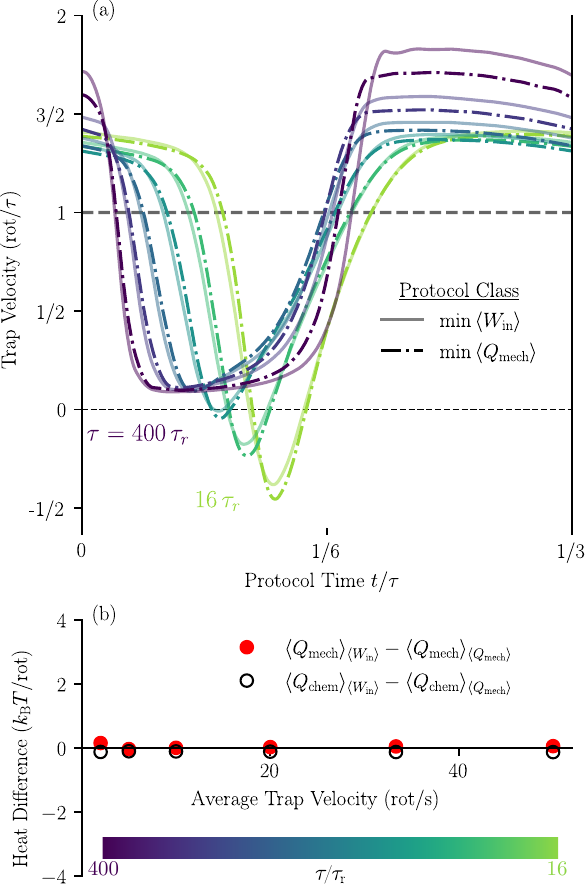}
    \caption{Results of designing protocols to reduce $\evsm{Q_\mathrm{mech}}$ instead of $\evsm{W_\mathrm{in}}$. Protocol durations are (across the colormap from purple to green): $400~\tau_r$, $160~\tau_r$, $80~\tau_r$, $40~\tau_r$, $24~\tau_r$, and $16~\tau_r$. (a) Trap velocity vs. protocol time for input-work-designed (solid curves) and mechanical-heat-designed (dash-dotted curves) protocols. The driving strategies are very similar. (b) Difference in heat production via the mechanical (red filled circles) and chemical (black open circles) channels between driving with the $\evsm{W_\mathrm{in}}$-optimized or $\evsm{Q_\mathrm{mech}}$-optimized protocols. The mechanical-heat-optimized protocol does reduce $\evsm{Q_\mathrm{mech}}$, at the expense of a larger increase in $\evsm{Q_\mathrm{chem}}$; however, the scale of these differences is small.}
    \label{fig:heat-opt}
\end{figure}

\section{Comparison to Langevin Simulations} \label{app:langevin}

To test our numerical method for time-evolution of the Fokker-Planck equation including switching~\eqref{eq:FPE_switching}, we compared the mean work predicted by our Fokker-Planck simulations to the mean work predicted from an ensemble of Langevin simulations under the same driving protocols.
We simulated \eqref{eq:langevin} in the same manner as Refs.~\cite{guptaOptimalControlF1ATPase2022,warehamMultiparameterOptimalControl2025}, with the landscape potential $\Uland^n$ making chemical transitions via a Poisson process with rates given by \eqref{eq:switching_rates}.

Figure~\ref{fig:langevin-compare} shows the work during our JAX-accelerated Fokker-Planck simulations and our Langevin simulations under the naive, linear-response-designed and AD-designed protocols.
We find broad agreement between the results of the two simulation methods, which is an indication that they simulate the same dynamics.

\section{Minimizing Mechanical Heat} \label{app:mech-minimizer}

To gauge the extent (if any) that our AD-designed protocols for switching landscapes sacrifice optimization of the mechanical heat to reduce the chemical heat, we performed the same optimization procedure with $\evsm{Q_\mathrm{mech}}$ (instead of $\evsm{W_\mathrm{in}}$) as the cost function.
Figure~\ref{fig:heat-opt}(a) shows trap velocity as a function of protocol time for varying protocol durations and both protocol-optimization classes, $\evsm{W_\mathrm{in}}$-optimized and $\evsm{Q_\mathrm{mech}}$.
The mechanical-heat-minimizing protocols are very similar to the work-minimizing protocols, and therefore have similar overall performance.
Figure~\ref{fig:heat-opt}(b) shows the difference in mechanical and chemical heat produced between driving with the two protocol-optimization classes.
The mechanical-heat-designed protocol almost universally improves over the work-designed protocol with respect to its optimization objective $\evsm{Q_\mathrm{mech}}$, but incurs a larger penalty in the chemical heat $\evsm{Q_\mathrm{chem}}$; however, these differences are small, about the order of $0.1~\kB T$ (much smaller than the additional heat production beyond $\evsm{Q_\mathrm{OT}}$ of order $10~\kB T$).
We conclude that in this case, most of the AD-designed protocols' reduction in the chemical heat $\evsm{Q_\mathrm{chem}}$ relative to the naive protocol can arise as a byproduct of reducing the mechanical heat $\evsm{Q_\mathrm{mech}}$; only a small fraction of improvement in the latter is sacrificed for the former. 

\bibliography{bibli}

\end{document}